# Assertion checker for the C programming language based on computations over event traces

**Mikhail Auguston**
Computer Science Department, New Mexico State University
Las Cruces, NM 88003, USA
phone: (505) 646-5286, fax: (505) 646-1002
mikau@cs.nmsu.edu
http://www.cs.nmsu.edu/~mikau

**ABSTRACT**

This paper suggests an approach to the development of software testing and debugging automation tools based on precise program behavior models. The program behavior model is defined as a set of events (event trace) with two basic binary relations over events -- precedence and inclusion, and represents the temporal relationship between actions. A language for the computations over event traces is developed that provides a basis for assertion checking, debugging queries, execution profiles, and performance measurements.

The approach is nondestructive, since assertion texts are separated from the target program source code and can be maintained independently. Assertions can capture the dynamic properties of a particular target program and can formalize the general knowledge of typical bugs and debugging strategies. An event grammar provides a sound basis for assertion language implementation via target program automatic instrumentation.

An implementation architecture and preliminary experiments with a prototype assertion checker for the C programming language are discussed.

**Keywords**

Program behavior models, events, event grammars, software testing and debugging automation.

## 1  INTRODUCTION

Program testing and debugging is still a human activity performed largely without any adequate tools, and consuming more than 50% of the total program development time and effort [9]. Testing and debugging are mostly concerned with the program run-time behavior, and developing a precise *model of program behavior* becomes the first step towards any dynamic analysis automation. In building such a model several considerations were taken in account. The first assumption we make is that the model is discrete, i.e. comprises a finite number of well-separated elements. For this reason the notion of *event* as an elementary unit of action is an appropriate basis for building the whole model. The event is an abstraction for any detectable action performed during the program execution, such as a statement execution, expression evaluation, procedure call, sending and receiving a message, etc.

Actions (or events) are evolving in time and the program behavior represents the temporal relationship between actions. This implies the necessity to introduce an ordering relation for events. Semantics of parallel programming languages and even some sequential languages (such as C) do not require the total ordering of actions, so *partial event ordering* is the most adequate method for this purpose [21].

Actions performed during the program execution are at different levels of granularity, some of them include other actions, e.g. a subroutine call event contains statement execution events. This consideration brings to our model *inclusion relation*. Under this relationship, events can be hierarchical objects and it becomes possible to consider program behavior at appropriate levels of granularity.

Finally, the program execution can be modeled as a set of events (*event trace*) with two basic relations: partial ordering and inclusion. In order to specify meaningful program behavior properties we have to enrich events with some attributes.

An event may have a type and some other attributes, such as event duration, program source code related to the event, program state associated with the event (i.e. program variable values at the beginning and at the end of the event), etc. This program behavior model may be regarded as a "lightweight" semantics of the programming language.

The next problem to be addressed after the program behavior model is set up is the formalism for specifying properties of the program behavior. This could be done in many different ways, e.g., by adopting some kind of logic calculi (predicate logic, temporal logic). Such a direction leads to tools for static program verification, or in more pragmatic incarnations to an approach called model checking [12].

Since our goal is dynamic program analysis that requires different types of assertion checking, debugging queries, program execution profiles, and so on, we developed the concept of a *computation over the event trace*. It seems that this concept is general enough to cover all the above mentioned needs in the unifying framework, and provides sufficient flexibility. This approach implies the design of a special programming language for computations over the event traces. We suggest a particular language called FORMAN ([3], [17]) based on a functional paradigm and the use of event patterns and aggregate operations over events. The papers [2], [3], [17] are based on our assertion checker prototype for a subset of the PASCAL language. This paper describes the first experience with an assertion checker for the C programming language. The implementation of the C assertion checker is based on source code automatic instrumentation and supports almost complete C language (the most serious constraint is the requirement that the target program is contained in a single compilation unit). To adjust to the specifics of the C target language the FORMAN language has been modified, in particular, the scope construct (WITHIN function-name) and explicit type cast have been added (see examples in Sec. 4).

Patterns describe the structure of events with context conditions. Program paths can be described by path expressions over events. All this makes it possible to write assertions not only about variable values at program points but also about data flow and control flow in the target program. Assertions can also be used as conditions in rules which describe debugging actions. For example, an error message is a typical action for a debugger or consistency checker. Thus, it is also possible to specify debugging strategies.

The notions of event and event type are powerful abstractions which make it possible to write assertions independent of a particular target program. Such generic assertions can be collected in standard libraries which represent general knowledge about typical bugs and debugging strategies and could be designed and distributed as special software tools.

Possible applications of a language for computations over a program event trace include program testing and debugging, performance measurement and modeling, program profiling, program animation, program maintenance and program documentation [5]. Even the traditional debugging method based on scattering print statements across the source code may be easily implemented as an appropriate computation on the event trace (see example in Sec 4). The advantage is that the print statements are kept in a separate file and the source code of the target program will be instrumented automatically just before execution. A study of applying FORMAN to parallel programming is presented in [4].

## 2 EVENTS

FORMAN is based on a semantic model of target program behavior in which the program execution is represented by a set of events. An *event* occurs when some action is performed during the program execution process. For instance, a function is called, a statement is executed, or some expression is evaluated. A particular action may be performed many times, but every execution of an action is denoted by a unique event.

Every event defines a time interval which has a beginning and an end. For atomic events, the beginning and end points of the time interval will be the same. All events used for assertion checking and other computations over event traces must be detectable by some implementation (e.g. by an appropriate target program instrumentation.) Attributes attached to events bring additional information about event context, such as current variable and expression values.

In order to give some rationale for our notion of an event, let us consider a well-known idea such as a counter. Usually the history of a variable X when used as a counter looks like:

X := 0; ...

Loop ...

X := X + 1; ...

endloop; ...

In order to determine whether the actual behavior of the counter X matches the pattern described by the program fragment above we have to consider the following events. Let Initialize_X denotes the event of assigning 0 to the variable X, Augment_X denotes the event of incrementing X, and Assign_X denotes the event of assigning any value to the variable X. The event Assign_X is a composite one; it contains either Initialize_X or Augment_X events. One could determine if X behaves as a counter when a program segment S is executed in the following way. First, the sequence A of all events of the type Assign_X from the event trace of program segment S has to be extracted preserving the ordering between events. Second, A has to be matched with the pattern:

```
Initialize_X (Augment_X)*
```

where '*' denotes repetition zero or more times. If the actual sequence of events does not match this pattern we can report an error. Therefore, assertion checking can be represented as a kind of computation over a target program event trace.

The program state (current values of variables) can be considered at the beginning or at the end of an appropriate event. This provides the opportunity to write assertions about program variable values at different points in the program execution history.

Program profiling usually is based on counting the number of events of some type, e.g. the number of statement executions or procedure calls. Performance measurements may be based on attaching the duration attribute to such events and summarizing durations of selected events.

## 3 PROGRAM BEHAVIOR MODEL

FORMAN is intended to be used to specify behavior of programs written in some high-level programming language which is called the *target language*. The model of target program behavior is formally defined as a set of events (*event trace*) with two basic relations, which may or may not hold between two arbitrary events. The events may be sequentially ordered (PRECEDES), or one of them might be included in another composite event (IN). For each pair of events in the event trace no more than one of these relations can be established.

In order to define the behavior model for a particular target language, types of events are introduced. Each event belongs to one or more of predefined event types, which are induced by target language abstract syntax (e.g. execute-statement, send-message, receive-message) or by target language semantics (e.g., rendezvous, wait, put-message-in-queue).

The target program execution model is defined by an event grammar. The event may be a compound object, in which case the grammar describes how the event is split into other event sequences or sets. The event grammar is a set of axioms that describe possible patterns of basic relations between events of different types in the program execution history; it is not intended to be used for parsing an actual event trace.

The rule `A :: B C` establishes that if an event `a` of the type A occurs in the trace of a program, it is necessary that events `b` and `c` of types B and C also exist, such that the relations `b IN a, c IN a, b PRECEDES c` hold.

For the C language assertion checker prototype we have defined the following simple event grammar.

(Axiom 1)   execute_program::

   ( ex_stmt | eval_expr )*

(Axiom 2)   ex_stmt::

   ( ex_stmt | eval_expr )*

(Axiom 3) eval_expr:: func_call |

   eval_expr+ destination? |

   { eval_expr } +

(Axiom 4) func_call::

   { eval_expr }* ex_stmt*

Axiom 1 states that the program execution event contains (the IN relation) a set of zero or more ordered (w.r.t. relation PRECEDES) events of the types execute-statement or evaluate-expression.

Axiom 2 states the same fact about the execute_statement event. For example, the event of executing a composite statement such as if-then-else will contain an event `eval_expr` for condition evaluation and a sequence of zero or more events for the corresponding THEN or ELSE branch execution. If a statement has a label attached, the label traversal itself is considered as an empty statement execution event.

Axiom 3 describes the possible structure of an expression evaluation event: it may contain a function call event or may be an ordered sequence of other expression evaluation events (e.g. for a 'comma" expression). The assignment expression evaluation contains the event `destination` which is distinguished because it is of a special importance for assertion checking. In our model we have avoided any assumptions about the ordering of argument evaluation for binary operations, such as '+' or '*', since the C language semantics leaves this undefined [18]. The metaexpression `{eval_expr}+` denotes a set of one or more events of the type `eval_expr` without any ordering relationship.

Axiom 4 describes the structure of a function call event which starts with a set (may be empty) of unordered events for actual parameter evaluation followed by the function body execution events.

The order of event occurrences reflects the semantics of the target language. When performing an assignment statement, first the right-hand part is evaluated and after this the destination event occurs (which denotes the assignment event itself). The event grammar makes FORMAN suitable for automatic source code instrumentation to detect all necessary events.

An event has attributes, such as the source text fragment from the corresponding target program, current values of target program variables and expressions at the beginning and at the end of event, the duration of the event, a previous path (i.e. set of events preceding the event in the target program execution history), etc.

FORMAN supplies a means for writing assertions about events and event sequences and sets. These include quantifiers and other aggregate operations over events, e.g., sequence, bag and set constructors, boolean operations and operations of the target language to write assertions about target program variables.

Events can be described by patterns which capture the structure of event and context conditions. Program paths can be described by regular path expressions over events.

## 4 EXAMPLES OF DEBUGGING RULES

In general, a *debugging rule* performs some actions that may include computations over the target program event trace. The aim is to generate informative messages and to provide the user with some values obtained from the trace in order to detect and localize bugs. Rules can provide dialog to the user as well. An assertion is a boolean expression that may contain quantifiers and sequencing constraints over events.

Assertions can be used as conditions in the rules describing actions that can be performed if an assertion is satisfied or violated. A debugging rule has the form:

```
assertion    SAY (expression sequence)
             ONFAIL SAY (expression sequence)
```

The presence of metavariables in the assertion makes it possible to use FORMAN as a debugger's query language. The evaluation of an assertion is interrupted when it becomes clear that the final value will be False (or True), and the current values of metavariables can be used to generate readable and informative messages.

We will use as an example of a C program the Simple Tokenizer program described in [25]. This program reads a text file until the special symbol '.' (dot) is read, recognizes small integers, identifiers, and some predefined key words, skips spaces and PASCAL-like comments, prints the input text with line numbers attached before each line, splits the output into pages with a page header on the top of each page (including page number), and reports each token recognized. Unrecognized symbols are printed as ERROR tokens. The source code contains 542 lines of code (including some of our updates and comments). The following list of function prototypes used in the Simple Tokenizer gives some idea of the architecture.

```
void init_scanner(char *name);
void init_page_header(char *name);
BOOLEAN get_source_line();
void get_char();
void skip_blanks();
void skip_comment();
void get_token();
void get_word();
BOOLEAN is_reserved_word();
void get_number();
void get_special();
void open_source_file(char *name);
void close_source_file();
void print_line(char line[]);
void print_token();
void print_page_header();
void quit_scanner();
```

The input text file for Simple Tokenizer used for running the following examples contained 150 lines of text with a total of 454 tokens. The corresponding output contained 13 pages with maximum of 50 lines per page (including the input lines and messages about tokens recognized, each on a separate line of output).

**Example of a debugging query.**

In order to obtain the history of a global variable `page_number` the following computation over the event trace can be performed. The WITHIN construct indicates the scope of the trace computations defined by this rule. The rule condition is TRUE, and as a side effect the entire history of variable `page_number` is shown. The [ ... ] list constructor defines a loop over the entire program event trace (`execute_program` event). All events matching the pattern `func_call IS printf` (i.e. events of the type func_call and function name 'printf') executed within the body of `print_page_header` function are selected from the trace and the function VALUE is applied to them. The metavariable C holds the event `func_call` under consideration. The resulting sequence consists of variable `page_number` values at the end of each event captured by metavariable C during the program execution.

```
WITHIN print_page_header
    TRUE
 SAY( 'The history of page_number variable
values is: '
 [ C: func_call IS 'printf'
    FROM execute_program
      APPLY VALUE(int)(AT C page_number) ]);
END
```

When executed on our prototype the following output is produced:

```
The history of page_number variable values
is: 1 2 3 4 5 6 7 8 9 10 11 12 13
```

This debugging rule provides a slice of the program execution history containing the trace of particular variable values. The matter of interest may be, for instance, to check whether the values in the variable history are arranged in ascending order.

**Example of an assertion checking.**

Let us write and check the assertion: *"There exists an input line with length exceeding some maximum, say 10."* The program snippet containing the function get_source_line

looks like:

```
BOOLEAN    get_source_line()
{char
print_buffer[MAX_SOURCE_LINE_LENGTH+9];
    if((fgets(source_buffer,
         MAX_SOURCE_LINE_LENGTH,
         source_file)) != NULL) {
    ++line_number;
Get_Line:
    sprintf(print_buffer, "%4d %d: %s",
    line_number,level,source_buffer);
    print_line(print_buffer);
    return(TRUE);
    }
    else return(FALSE);          }
```

Traversal of a label is an event of the type `ex_stmt`, and we can check the value of a C expression `strlen(source_buffer) > 10` after this event.

```
WITHIN get_source_line
    EXISTS L: ex_stmt IS 'Get_Line:'
         FROM execute_program
VALUE(int)(AT L strlen(source_buffer) >10)
SAY('Too long input line detected at stmt' )
SAY(L)
SAY( 'It is '
    VALUE(int)(AT L strlen(source_buffer))
     'characters long')
ONFAIL SAY(' No long input lines detected');
```

We check whether the expression `strlen(source_buffer) > 10` is not equal to 0 for all events L. When the assertion is satisfied for the first time, the assertion evaluation terminates and the current value of the metavariable L can be used for message output. In order to make error messages more informative, the value of a metavariable when printed by the SAY clause is shown in the form:

```
event-type:> event-source-text
source_line_number within function_name
Time= event-begin-time .. event-end-time
```

Event begin and end times in this prototype implementation are simply values of the step counter.

When executed on our prototype this assertion checking yields the following output.

```
Too long input line detected at stmt
ex_stmt :> 'Get_Line:' source line 460 within function get_source_line
Time= 95 .. 96
It is 20 characters long
```

**Example of a run time statistics gathering.**

It is hard to measure real execution time of a heavily instrumented target program, although the simulated time measurement may be performed given that events may have some duration attributes predefined. In order to obtain the actual number of function calls executed, number of function `get_source_line` calls, and number of tokens recognized by the Simple Tokenizer, the following query can be performed:

```
TRUE
SAY('Total function calls'
    CARD[ ALL func_call
         FROM execute_program])
SAY('Total function get_source_line calls'
    CARD [ func_call IS get_source_line
         FROM execute_program])
SAY('Total tokens recognized'
    CARD [ ALL func_call IS get_token
         FROM execute_program]
    ', among them '
    CARD [ ALL F: func_call &
     SOURCE_TEXT(F) == 'get_token'
     AND VALUE (int)(AT F token == ERROR)
            FROM execute_program]
    'ERROR tokens detected' );
```

The CARD operator returns the number of items selected by the aggregate operation, i.e. the number of events matching the pattern in the aggregate operation body. The ALL option in the aggregate operation indicates that all nested events of the type `func_call` should be taken into account. The pattern in the third aggregate operation provides an example of a complex event pattern with a context condition attached. The scope of this trace computation is the entire program trace. After execution on our prototype the following output is obtained.

```
Total function calls 6802
Total function get_source_line calls 150
Total tokens recognized 454, among them 37
    ERROR tokens detected
```

**Example of path expression checking.**

Regular expressions over event patterns may describe sequences of events extracted from the event trace. The following assertion checks whether function get_token and print_token calls appear in a certain order. Sequence of events satisfying the pattern X:func_call& SOURCE_TEXT(X) == 'get_token' OR SOURCE_TEXT(X)=='print_token' is selected from the entire event trace and matched against the path expression (func_call IS 'get_token' func_call IS 'print_token') +. A message is produced with information about the pattern matching results.

```
[ X: func_call & SOURCE_TEXT(X)==
            'get_token' OR
                  SOURCE_TEXT(X)==
'print_token' FROM execute_program ]
     SATISFIES(func_call IS 'get_token'
      func_call IS 'print_token' ) +
  SAY('function calls follow the pattern
     (get_token print_token) + ')
 ONFAIL SAY( 'pattern
     (get_token print_token) +
     is violated');
```

**Example of instrumenting the target source code with print statements.**

Suppose we want to insert in the target source code print statements to print at run time the value of input strings with length exceeding 10 and corresponding line numbers. Values of interest are available in global variables source_buffer and line_number, respectively. The following debugging rule performs this function.

```
WITHIN get_source_line
FOREACH L1: ex_stmt IS 'Get_Line:'
        FROM execute_program
   VALUE ( int )
 ( AT  L1  strlen(source_buffer)>10?
printf("long line!!!\n%s\n",source_buffer):1)
   AND
       VALUE ( int )
  ( AT L1
   printf("line_number=%d\n",line_number));
 END
```

Formally this rule will cause an assertion checking, which will be successful since the C expression involved yields a non-zero value (representing Boolean TRUE); as a side effect the print statements are executed at run time. This debugging rule has two aspects worthy of notice. First, the instrumentation code is separated from the target code; it will be inserted automatically just before the execution and can be maintained in a separate file. There may be several different print instrumentations defined for the same target program; keeping them in separate files provides a great flexibility in arranging a custom set of print statements to be inserted at run time. Second, the instrumentation is attached to a particular event in the trace matching the pattern ex_stmt IS 'Get_Line:', i.e. traversal of the label Get_Line:, therefore it does not depend on possible target code modifications as long as the label is not changed.

Debugging rules can be considered as a way of formalizing reasoning about the target program execution -- humans often use similar patterns for reasoning when debugging programs. For example, if the index expression of an array element is out of range, the debugger can try a rule for eval-index events that invokes another rule about a wrong value of the event eval-expression, which in turn will cause investigation of histories of all variables included in the expression.

## 5  BRIEF IMPLEMENTATION SURVEY

The architecture of the computations over the event traces for the C programming language is based on the automatic instrumentation of the target program source code in such a way that some computations over the trace are performed at run time and the rest of information is saved in the trace file for postmortem processing. The instrumentation does not change the semantics of the target program. The trace file is read by the FORMAN interpreter to complete the computations over the trace and to generate messages. A special attempt in this prototype was made to optimize the trace generation, in particular to filter events in order to reduce the size the trace.

The front end of the assertion checker was adapted and modified from Shawn's Flisakowski parser and abstract syntax tree builder for the complete C programming language (gcc version) [14]. The instrumentation module was designed by Ana Erendira Flores-Mendoza as her Master's project in the NMSU CS Department [15]. The total size of the software used for the prototype amounts to more then 20KLOC of C/

lex/yacc/Rigal [1] code.

Since an event in our model has a duration and may contain another events, it is represented on the trace by two records, one for the beginning of event and one for the end. The semantics of the C language do not specify the order of subexpression execution; to address this issue and to ensure proper nesting of event eval_expr beginning and end records on the trace the instrumented code maintains some auxiliary stack for expression evaluation. A similar stack mechanism is added to the instrumented code to maintain proper nesting of ex_stmt and func_call events when performing return, goto, and break statements. These specifics of our target program behavior model led as to the decision to implement the instrumentation module from the scratch rather than to use some generic instrumentation tools like [33]. The basic building block for expression E instrumentation is comma-expression (e1, temp = E, e2, temp), where e1 stands for prologue instrumentation, e2 stands for epilog instrumentation, and temp variable holds the result of the original expression E evaluation.

Only events necessary for the given FORMAN program are involved in the computations over the trace and put on the trace. For the Simple Tokenizer program discussed above, using the input file with 150 lines and 454 tokens and the entire set of debugging rules described in the previous section the total number of events generated by the target program according to the event grammar is 105,808, although only 7253 of them (less then 7%) are put on the trace. Even in its current state with many potential optimizations not yet implemented, the prototype demonstrates the feasibility of trace computations for "typical" student programs like the Simple Tokenizer. Our experiments with other C programs show that storing several tens of thousands of events on the trace is sufficient for a large number of "typical" C programs run with a set of debugging rules and assertions similar to the examples in Sec. 4. It should be noted that typically the size of input data used for testing and debugging purposes is relatively small.

## 6 RELATED WORK

What follows is a very brief survey of basic ideas known in Debugging Automation to provide the background for the approach advocated in this paper.

**Event Notion**

The Event Based Behavioral Abstraction (EBBA) method suggested in [7] characterizes the behavior of the entire program in terms of both primitive and composite events. Context conditions involving event attribute values can be used to distinguish events. EBBA defines two higher-level means for modeling system behavior -- clustering and filtering. Clustering is used to express behavior as composite events, i.e. aggregates of previously defined events. Filtering serves to eliminate from consideration events which are not relevant to the model being investigated. Both event recognition and filtering can be performed at run-time.

An event-based debugger for the C programming language called Dalek [27] provides a means for describing user-defined events which typically are points within a program execution trace. A target program has to be instrumented in order to collect values of event attributes. Composite events can be recognized at run-time as collections of primitive events.

FORMAN has a more comprehensive modeling approach than EBBA or Dalek, based on the event grammar. A language for expressing computations over execution histories is provided, which is missing in EBBA and Dalek. The event grammar makes FORMAN suitable for automatic source code instrumentation to detect all necessary events. FORMAN supports the design of universal assertions and debugging rules that could be used for debugging of arbitrary target programs. This generality is missing in the EBBA and Dalek approaches. The event in FORMAN is a time interval, in contrast with the event notion in previous approaches where events are considered pointwise time moments.

The COCA debugger [13] for the C language uses the GDB debugger for tracing and PROLOG for debugging queries execution. It provides a certain event grammar for C traces and event patterns based on attributes for event search. The query language is designed around special primitives built into the PROLOG query evaluator. We assume that FORMAN is more suitable for trace computations as it has been designed for this specific purpose.

**Path Expressions**

Data and control flow descriptions of the target program are essential for testing and debugging purposes. It is useful to give such a description in an explicit and precise form. The path expression technique introduced for specifying parallel programs in [11] is one such formalism. Trace specifications also are used in [26] for software specification. This technique has been used in several projects as a background for high-level debugging tools, (e.g. in [10]), where path rules are suggested as a kind of debugger commands. FORMAN provides a flexible language means for trace specification including event patterns and regular expressions over them.

**Assertion Languages**

Assertion (or annotation) languages provide yet another approach to debugging automation. The approaches currently in use are mostly based on boolean expressions attached to selected points of the target program, like the assert macro in C [18]. The ANNA [23] annotation language for the Ada target language supports assertions on variable and type declarations. In the TSL [22], [29] annotation language for Ada the notion of event is introduced in order to describe the behavior of Tasks. Patterns can be written which involve parameter values of Task entry calls. Assertions are written in Ada itself, using a number of special pre-defined predicates. Assertion-checking is dynamic at run-time, and does not need post-mortem analysis. The RAPIDE project [24] provides an event-based assertion language for software architecture description.

In [6] events are introduced to describe process communication, termination, and connection and detachment of process to channels. A language of Behavior Expressions (BE) is provided to write assertions about sequences of process interactions. BE is able to describe allowed sequences of events as well as some predicates defined on the values of the variables of processes. Event types are process communication and interactions such as send, receive, terminate, connect, detach. Evaluation of assertions is done at run-time. No composite events are provided.

Another experimental debugging tool is based on trace analysis with respect to assertions in temporal interval logic. This work is presented in [20] where four types of events are introduced: assignment to variables, reaching a label, interprocess communication and process instantiation or termination. Composite events cannot be defined. Different varieties of temporal logic languages are used for program static analysis called Model Checking [12].

In [30] a practical approach to programming with assertions for the C language is advocated, and it is demonstrated that even local assertions associated with particular points within the program may be extremely useful for program debugging.

The DUEL [19] debugging language introduces expressions for C aggregate data exploration, for both assertions and queries.

The FORMAN language for computations over traces provides a flexible means for writing both local and global assertions, including those about temporal relations between events.

**Algorithmic Debugging**

The original algorithmic program debugging method was introduced in [32] for the Prolog language. In [31] and [16] this paradigm is applied to a subset of PASCAL. The debugger executes the program and builds a trace execution tree at the procedure level while saving some useful trace information such as procedure names and input/output parameter values. The algorithmic debugger traverses the execution tree and interacts with the user by asking about the intended behavior of each procedure. The user has the possibility to answer "yes" or "no" about the intended behavior of the procedure. The search finally ends and a bug is localized within a procedure $p$ when one of the following holds: procedure $p$ contains no procedure calls, or all procedure calls performed from the body of procedure $p$ fulfill the user's expectations.

Algorithmic debugging can be considered as an example of debugging strategy, based on some assertion language (in this case assertions about results of a procedure call). The notion of computation over execution trace introduced in FORMAN may be a convenient basis for describing such debugging strategies.

# 7 CONCLUSIONS

In brief, our approach can be explained as "computations over a target program event trace based on a precise program behavior model". According to [8] and [28], approximately 40-50% of all bugs detected during the program testing are logic, structural, and functionality bugs, i.e., bugs which could be detected by appropriate assertion checking similar to that demonstrated above.

We expect the advantages of our approach to be the following:

- The notion of **an event grammar** provides a general basis for program behavior models. In contrast with previous approaches, the **event** is not a point in the trace but an interval with a beginning and an end.
- Event grammar provides a coordinate system to refer to any interesting event in the execution history. Event attributes provide complete **access to each target program's execution state**. Assertions about particular execution states as well as assertions about sets of different execution states may be checked.
- The IN relation yields a **hierarchy of events**, so the assertions can be defined at an appropriate level of granularity.
- A language for **computations over event traces** provides a **uniform framework** for assertion checking, profiles, debugging queries, and performance measurements.
- The fact that assertions and other computations over the target program event trace can be **separated from the text of the target program** allows accumulation of formalized knowledge about particular programs and makes it easy to control the number of assertions to be checked.

The first experiments with our C assertion checker prototype prove that:

- instrumentation of the C source code may be an appropriate technique for automatic testing and debugging tool design,
- event filtering can reduce the size of the stored event trace to 5-20% of the total trace,
- the size of the stored event trace could be kept within reasonable limits (several tens of thousands of events) for realistic C programs.

The future work will be dedicated to further optimizations of trace computation and event filtering, and to the design of an appropriate user interface.

**ACKNOWLEDGEMENTS**

I would like to thank Jonathan Cook, Larry King, and Hue McCoy for valuable remarks and suggestions.

This work was supported in part by NSF grant #9810732.